\documentclass[prd,amsmath,notitlepage,superscriptaddress,twocolumn,nofootinbib]{revtex4-2}

\pdfoutput=1
\usepackage{graphicx,ulem}
\usepackage{float}
\usepackage{tensor}
\usepackage{epstopdf,cancel}
\usepackage{epsf,latexsym,bbm,euscript}
\usepackage{amssymb,amsmath}
\usepackage{mathtools}
\usepackage{times,graphics}
\usepackage{soul,xcolor}
\usepackage{mathtools}

\usepackage{enumitem}

\usepackage{xcolor}

\usepackage{mathtools}

\makeatletter
\newcommand*{\defeq}{\mathrel{\rlap{%
	\raisebox{0.3ex}{$\m@th\cdot$}}%
	\raisebox{-0.3ex}{$\m@th\cdot$}}%
	=}
\newcommand*{\eqdef}{=\mathrel{\rlap{%
	\raisebox{0.3ex}{$\m@th\cdot$}}%
	\raisebox{-0.3ex}{$\m@th\cdot$}}%
	}
\makeatother
		
\newcommand{\sg}{\textsl{g}}
\newcommand{\ah}{\mathrm{AH}}
\def\6{{\langle}}
\def\9{{\rangle}}
\def\pad{{\partial}}
\newcommand{\be}{\begin{equation}}
\newcommand{\ee}{\end{equation}}

\def\bt{{\bar{t}}}
\def\bC{{\bar{C}}}
\def\PG{{\mathrm{PG}}}

\usepackage{url,hyperref}
\hypersetup{colorlinks,linkcolor={blue!55!black},citecolor={red!45!black},urlcolor={blue!45!black},breaklinks=true}

\begin{document}

\title{Paradoxes before the paradox: surface gravity and the information loss problem}

\author {Robert B.\ Mann}
\email{rbmann@uwaterloo.ca}
\affiliation{Department of Physics and Astronomy, University of Waterloo, Waterloo, Ontario N2L 3G1, Canada}
\affiliation{Perimeter Institute for Theoretical Physics, Waterloo, Ontario N2L 6B9, Canada}

\author{Sebastian Murk}
\email{sebastian.murk@mq.edu.au}
\affiliation{Department of Physics and Astronomy, Macquarie University, Sydney, New South Wales 2109, Australia}
\affiliation{Sydney Quantum Academy, Sydney, New South Wales 2006, Australia}

\author{Daniel R.\ Terno}
\email{daniel.terno@mq.edu.au}
\affiliation{Department of Physics and Astronomy, Macquarie University, Sydney, New South Wales 2109, Australia}

\begin{abstract}
The information loss paradox is widely regarded as one of the biggest open problems in theoretical physics. Several classical and quantum features must be present to enable its formulation. First, an event horizon is needed to justify the objective status of tracing out degrees of freedom inside the black hole. Second, evaporation must be completed (or nearly completed) in  finite time according to a distant observer, and thus the formation of the black hole should also occur in finite time. In spherical symmetry these requirements constrain the possible metrics strongly enough to obtain a unique black hole formation scenario and match their parameters with the semiclassical results. However, the two principal generalizations of surface gravity, the quantity that determines the Hawking temperature, do not agree with each other on the dynamical background. Neither can correspond to the emission of nearly-thermal radiation. We infer from this that the information loss problem cannot be consistently posed in its standard form.
\end{abstract}

\maketitle

\section{Introduction}
Information loss in black hole evolution is one of the longest-running controversies in theoretical physics \cite{h:76,wald:01,rev-0,rev-00,rev-1,rev-12,rev-12marolf,coy:15,info:21,SFA:21,bfm:21}. Its essence is captured by the following scenario:  according to distant observers, matter collapsing into a black hole completely evaporates via Hawking radiation within a finite time. If quantum correlations between the inside and outside of the black hole horizon are not restored during the evaporation, this evolution of low-entropy collapsing matter into high-entropy radiation implies information loss. This problem is referred to as a paradox because a combination of information-preserving theories --- quantum field theory and general relativity (GR) --- ostensibly leads to a loss of information \cite{bmt:17}.

Its status as a paradox, the necessity and/or validity of particular resolutions and their implications for a putative theory of quantum gravity or the  fundamental structure of quantum theory are not the subject of our discussion here. Instead, we focus on the consequences of its formulation within the framework of semiclassical gravity. In common with the paradoxes of quantum mechanics, the information loss problem combines classical and quantum elements and some counterfactual reasoning. In this paper, we consider the physical and mathematical consequences of having the necessary elements for its formulation realized.

We find that the conditions required for the formulation of the paradox (in contrast to its resolution) cannot be realized without significant modifications of the late-time black hole radiation, which is considered to be one of the most established results of quantum field theory in curved spacetime. The key technical findings that we report are the discordant properties of generalizations of surface gravity. As a result, we conclude that, while gravitational collapse and gravitationally-induced radiation contain several important physical questions, including matter-gravity correlations, observability of various horizons, and the applicability of semiclassical physics, the standard formulation of apparent loss of information cannot  {consistently be made} in the context of semiclassical gravity. Consequently, if the paradox cannot be self-consistently formulated in the best tested framework we currently have available, this suggests that its various proposed resolutions should be reappraised.

We first note that the setting for the formulation of the information loss problem involves at least the following:
\begin{enumerate}
\item Formation of a transient trapped region. Such a region either completely disappears or turns into a stable remnant; in either case, this takes place  in finite time as measured by a distant observer Bob. This provides the scattering-like setting to describe the states (and their alleged information content) ``before'' and ``after''.
\item Formation of an event horizon (and not just any other special surface). Its existence is necessary to provide an objective, observer-independent separation of the spacetime into accessible and inaccessible regions, and it is only with respect to this boundary that tracing out of the interior degrees of freedom is not just a technical limitation (akin to our inability to recover correlations between the smoke and information that was contained in the proverbial burned encyclopedia), but a fundamental physical restriction \cite{wald:01,vis:08}.
\item Thermal or nearly-thermal character of the radiation. It is responsible for the eventual disappearance of the trapped region and for the high entropy of the reduced exterior density operator.
\end{enumerate}
Additional assumptions may or should be made to enable a particular formulation of the paradox, but the triad of finite lifetime, event horizon, and temperature are the ineluctable components of the paradox's formulation.

\begin{figure}[htbp!]
	\centering
	\includegraphics[width=0.45\textwidth]{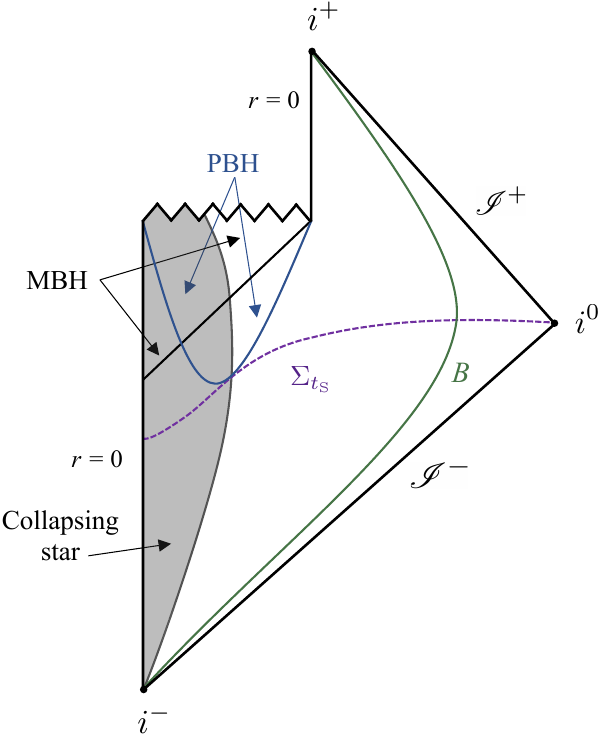}
	\caption{
		Schematic depiction of collapse into and complete evaporation of a black hole as outlined in Ref.~\cite{h:76}. Spacetime regions corresponding to physical black hole (PBH) and mathematical black hole (MBH) solutions are indicated by arrows. Different proposals for the spacetime structure after the singular corner point that corresponds to complete evaporation appear in various resolutions of the information loss paradox (see, e.g.\ Refs.~\cite{rev-0,rev-00,rev-1,coy:15,bfm:21}). Finite formation time $t_\mathrm{S}$ of the apparent horizon according to a distant observer is a necessary condition for the formulation of the paradox. Part of the equal time surface $\Sigma_{t_\mathrm{S}}$ is shown as a dashed purple line. The outer apparent  horizon $r_\sg(t)$  and the inner apparent horizon form the boundary of a PBH and are shown in blue. The apparent horizon $r_\sg(t)$ is a timelike hypersurface during its entire existence \cite{bmmt:19}. Bob's trajectory is indicated by the green curve. The collapsing matter and its surface are shown as in conventional depictions of the collapse. However, the matter in the vicinity of the outer apparent horizon  $\big(t,r_\sg(t)\big)$ violates the null energy condition (NEC) for $t \geqslant t_\mathrm{S}$. Moreover, the energy density, pressure, and flux as seen by an infalling observer Alice vary continuously across it, and the equation of state dramatically differs from that of normal matter that may have been used to model the initial energy-momentum tensor (EMT) of the collapse (see Sec.~\ref{horizon} and Ref.~\cite{mut:21} for details).
	}
	\label{schema1}
\end{figure}

The logical framework of our result is as follows: the existence of a transient trapped region implies its formation at some finite time $t_\mathrm{S}$ as measured by Bob (Sec.~\ref{sec:prerequisites}). Together with the minimal regularity assumption (finiteness of all curvature scalars that are obtained as polynomial invariants of the Riemann tensor at the apparent horizon), this constrains the possible spherically symmetric geometries enough to prescribe a unique formation scenario (Sec.~\ref{horizon}) that requires us to generalize the notion of surface gravity. Fig.~\ref{schema1} schematically represents the geometry that underpins the paradox.

Dynamical black hole spacetimes do not possess a timelike Killing field, and thus require different methods to define the surface gravity. The literature contains several possible definitions that  are broadly classified according to which of the (equivalent in the stationary case) properties of surface gravity they are related. The results serve as analogs of the Hawking temperature, which they approach in a suitable limit. Under quite general conditions these classes provide close values for their respective quantities. However, we will show that these conditions are not satisfied if the apparent horizon is formed at finite $t_\mathrm{S}$. Consequently, these values differ significantly, and none of them can approach the Hawking temperature  $1/4M$ without violating the semiclassical luminosity relation $L\propto M^{-2}$ as we shall demonstrate (Sec.~\ref{kappaT}).

Our article is organized as follows: in the next section we review the assumptions of semiclassical black hole physics. We restrict our discussion to spherical symmetry. Then, we translate the necessary requirements for the formulation of  the information loss problem into conditions on self-consistent solutions of the Einstein equations. In Sec.~\ref{horizon}, we summarize the properties of these solutions, emphasizing the near-horizon geometry and the unique scenario of black hole formation. In Sec.~\ref{idUx}, we identify the leading terms in the self-consistent metric using a general evaporation law. Sec.~\ref{kappaT} outlines the consequences of this identification; we demonstrate that the two natural candidates for the Hawking temperature, when evaluated for the configurations of Sec.~\ref{horizon}, disagree with each other and cannot be reconciled with the standard semiclassical result  without contradicting the results of Sec.~\ref{idUx}.

We use the $(-+++)$ signature of the metric and work in units where $\hbar=c=G=k_B=1$.  Derivatives of a function of a single variable are marked with a prime:  $r_\sg'(t)\equiv dr_\sg/dt$, $r_+'(v)\equiv dr_+/dv$, etc. Derivatives with respect to the proper time $\sigma$ are denoted by a dot, $\dot r=dr/d\sigma$.

\section{Prerequisites for the paradox} \label{sec:prerequisites}
We work in semiclassical gravity. That means we use classical notions (horizons, trajectories, etc.) and describe dynamics via the Einstein equations $G_{\mu\nu}=8\pi T_{\mu\nu}$, where the standard Einstein tensor on the left-hand side is equated to the expectation value of the renormalized energy-momentum tensor (EMT), $T_{\mu\nu} = \langle \hat T_{\mu\nu} \rangle_\omega$ \cite{hv:book,pp:09,bmt:18}. The quantum state $\omega$ represents both the collapsing matter and the created excitations of the quantum fields.

A general spherically symmetric metric in Schwarzschild coordinates is given by
\begin{align}
	ds^2=-e^{2h(t,r)}f(t,r)dt^2+f(t,r)^{-1}dr^2+r^2d\Omega, \label{sgenm}
\end{align}
where $r$ is the areal radius \cite{he:book,c:book}. The function $f(t,r)=1-C(t,r)/r$ is coordinate-independent. The Misner\textendash{}Sharp (MS) mass \cite{ms,faraoni:b,aphor} $C(t,r)/2$ is invariantly defined via
\begin{align}
	1-C/r \defeq \partial_\mu r \partial^\mu r \; . \label{defMS}
\end{align}
The same geometry can be described using the advanced null coordinate $v$ as
\begin{align}
  ds^2=-e^{2h_+}\left(1-\frac{C_+}{r}\right)dv^2+2e^{h_+}dvdr +r^2d\Omega. \label{lfv}
\end{align}
Invariance of the MS mass implies $C_+(v,r)=C\big(t(v,r),r\big)$, while the functions $h_+(v,r)$ and $h(t,r)$ are the integrating factors in various coordinate transformations, such as
\begin{align}
	dt = e^{-h} \big(e^{h_+}dv- f^{-1}dr\big). \label{intfu}
\end{align}
The study of null geodesics and their congruences is one of the principal tools of black hole physics. Assuming spherical symmetry, radial null geodesics are determined in Schwarzschild coordinates as the solutions of
\begin{align}
	\frac{dr}{dt} = \pm e^{h}f, \label{nullg}
\end{align}
where for $f>0$ the upper sign corresponds to an outgoing geodesic. In $(v,r)$ coordinates the ingoing geodesics correspond to $v=\mathrm{const}$, and the outgoing geodesics satisfy
\begin{align}
	\frac{dr}{dv} = {e^{h_+}f}. \label{eq:vr-out-geod}
\end{align}
We assume that the spacetime is asymptotically flat and $t$ is the physical time of a distant observer (Bob) to simplify the exposition, though we emphasize that it is not necessary to assume any particular structure at infinity to derive the results we present in the next section.

A future event horizon \cite{he:book,c:book,faraoni:b,fn:book} is a causal boundary separating the domain of outer communication (the region from which it is possible to send signals to any future asymptotic observer) from the rest of spacetime (regions where this is not possible). However, to determine its existence/presence requires knowledge of the entire history of spacetime (and therefore also infinitely far into its future) \cite{fn:book,faraoni:b,vis:08,mV:2014}. In what follows we refer to the causally disconnected spacetime domain as a {\it mathematical black hole} (MBH) \cite{c:book,frolov-def}.

A much more practical and useful definition captures the idea of a black hole as part of space from which nothing can escape at a given moment in time. A trapped region is a domain where both ingoing and outgoing future-directed null geodesics emanating from a spacelike two-dimensional surface with spherical topology have negative expansion. The apparent horizon \cite{he:book,faraoni:b,aphor} is its evolving outer boundary. In general this notion depends on the spacetime foliation, but the apparent horizon is unambiguously defined in all foliations that respect spherical symmetry \cite{faraoni:b,aphor}.

In $(v,r)$ coordinates the expansions of ingoing and outgoing radial geodesic congruences with the tangents
\begin{align}
	n^\mu = \big(0,-e^{-h_+},0,0\big),  \quad l^\mu = \big(1,\tfrac{1}{2} e^{h_+} f,0,0\big),
\end{align}
that satisfy $n\cdot l=-1$ are
\begin{align}
	\theta_n = - \frac{2e^{-h_+}}{r}, \qquad \theta_l = \frac{e^{h_+} f}{r},
\end{align}
respectively. The apparent horizon is located at the Schwarzschild radius $r_\sg(t) \equiv r_+(v)$, namely the largest root of $f(t,r)=0$. Following the nomenclature of Ref.~\cite{frolov-def}, we refer to its interior as a {\it physical black hole} (PBH).

Using the retarded null coordinate $u$ leads to the metric in the form
\begin{align}
	ds^2=-e^{2h_-}\left(1-\frac{C_-}{r}\right)du^2-2e^{h_-}dudr +r^2d\Omega, \label{lfu}
\end{align}
which is particularly suitable for describing the spacetime of a white hole (then the Schwarzschild radius $r_-(u)$ is the boundary of the anti-trapped region where both expansions are positive).

In classical GR the event and the apparent horizon are regular surfaces: the curvature scalars, such as the Ricci curvature $R$ and the Kretschmann scalar $R^{\mu\nu\rho\sigma}R_{\mu\nu\rho\sigma}$ are finite. In studies of field theories on curved backgrounds this assumption is necessary to maintain predictability of the theory \cite{hv:book,pt:book}.

The event horizon is an indispensable concept in formulating the paradox \cite{wald:01,bmt:17,vis:08}. Tracing out the inaccessible degrees of freedom naturally leads to entropy production in an overall unitary evolution. To differ from non-paradoxical entropy increases common in thermodynamic subsystems, this separation of spacetime regions should not represent a practical limitation on a distant Bob, but rather an absolute physical restriction. This is provided by an event horizon that bounds the absolutely inaccessible spacetime region according to Bob, which is distinct from the transient (albeit extremely long-lived) trapped region of regular black holes \cite{bar:68,fv:81,hay:06,f:14,cflpv:18,bm:19,cp:rev}.

According to Bob, the formation of black holes from classical collapsing matter takes an infinite amount of time. After at most a few dozen multiples of light-crossing time $r_\sg$, he cannot receive signals from an infalling observer Alice (an observer co-moving with the matter who is initially at or near its edge); the redshift requires that the energy of any such detected signal is greater than the mass of the black hole.

Consequently, in classical GR the event horizon and Alice's experiences, like crossing the Schwarzschild radius, are counterfactual \cite{bmt:17}. Her clock readings should indicate various processes occurring at finite proper times $\tau_i$. As Alice cannot communicate her clock readings to Bob, these are experimentally unverifiable consequences of the formalism of GR. Nevertheless, a finite proper crossing time promotes the event horizon, and by extension the quantum states associated with the black hole horizon and its interior, from convenient mathematical concepts to physical entities in the theory.

The paradox that is based on properties of Hawking-like radiation cannot be constructed in such a case. Even if the existence of collapse-induced radiation does not require a horizon for its production \cite{pp:09,haj:87,blsv:06,vsk:07} (thereby resolving an obvious causal difficulty of the collapse and evaporation process taking a finite time according to Bob), formation of the event horizon should also occur at some finite time $t_*$ that precedes the evaporation time $t_\mathrm{e}$.

If the null energy condition (NEC) \cite{mmv:17,ks:20} is satisfied, i.e.\ for any null vector $k^\mu$, $k^\mu k_\mu=0$, contraction with the EMT is non-negative, $T_{\mu\nu}k^\mu k^\nu \geqslant 0$, then the apparent horizon is located inside of the event horizon \cite{he:book,fn:book}. This condition is violated by Hawking radiation. A detailed semiclassical analysis subsequently indicates that part of the trapped region is outside of the MBH \cite{fn:book, faraoni:b,bardeen:81}.

For an evaporating black hole ($r'_\sg \defeq d r_\sg/dt<0$) a weaker statement --- existence of the event horizon in finite $t$ implies formation of the apparent horizon at some finite time $t_\mathrm{S}$ --- ensues on the following logical grounds: consider an outward-pointing radial null geodesic that is emitted from a location $(r,t)$ that is outside of the apparent horizon $r_\sg(t)$. Eq.~\eqref{nullg} indicates that $r'(t) =: v(t)>0$, and as $r_\sg'(t)<0$ this is true along the entire trajectory. For this geodesic to avoid reaching  infinity as $t\to\infty$,  at least either $\lim_{t\to\infty}\lim_{r\to r_*}h=-\infty$ or  $\lim_{t\to\infty}\lim_{r\to r_*}f=0$  should hold  for some $r_*<\infty$. The former is impossible as the Schwarzschild coordinates are regular outside of $r_\sg$,  while the latter contradicts the definition of the MS mass and its relationship with the apparent horizon as $r_\sg(t)$ is the largest root of $f=0$.  Hence such a geodesic does reach future null infinity, and a geodesic that is emitted outside of the receding apparent horizon is not contained within a MBH.

This discussion leads to two conditions that are necessary for the formulation of the information loss problem. First, an apparent horizon must be a regular surface to ensure predictability, and second, it must form at some finite time according to Bob (otherwise formation of the event horizon prior to evaporation of the black hole is impossible, thus preventing formulation of the alleged paradox). In spherical symmetry this is enough to describe the black hole formation scenario and geometry near the apparent horizon \cite{mut:21}.

\section{Near-horizon geometry} \label{horizon}
Both the regularity conditions and the Einstein equations can be conveniently expressed in terms of combined expressions
\begin{align}
	\tau_t \defeq e^{-2h} T_{tt}, \qquad \tau^r \defeq T^{rr}, \qquad \tau_t^{~r} \defeq e^{-h}T_t^{~r} ,
	\label{eq:effEMTtaus}
\end{align}
that are used instead of the EMT components \cite{bmmt:19}. In particular, the three Einstein equations for $G_{tt}$, $G_t^{~r}$, and $G^{rr}$ are
\begin{align}
	\partial_r C &= 8 \pi r^2 \tau_t / f , \label{gtt} \\
	\partial_t C &= 8 \pi r^2 e^h \tau_t^{~r} , \label{gtr} \\
	\partial_r h &= 4 \pi r \big(\tau_t + \tau^r\big)/f^2 , \label{grr}
\end{align}
respectively.

The requirement of regularity of the apparent horizon and existence of real solutions describing geometry in its vicinity constrain the generic limiting form of the EMT and, eventually, the metric in its vicinity. Regularity is expressed as the demand that curvature scalars obtained from polynomials of components of the Riemann tensor are finite. However, in practice it is sufficient to require that the contractions $T^{\mu\nu}T_{\mu\nu}$ and $T^\mu_{~\mu}$ are finite at the apparent horizon to satisfy the regularity requirement \cite{t:19}. Leading terms in the reduced EMT components can in principle scale as $\tau_a\propto f^{k_a}$, for some powers $k_a$ with $\tau_a$ being one of $\tau_{t}$, $\tau^{r}$, and $\tau_t^{~r}$.

However, a careful analysis shows that only two solutions are possible: those that satisfy $k_a\equiv k=0$ and a subset of the solutions with $k_a\equiv k=1$ \cite{mut:21,t:19,t:20}. In the former case the metric functions that solve Eqs.~\eqref{gtt} and~\eqref{grr} are
\begin{align}
	C= r_\sg - 4 \sqrt{\pi} r_\sg^{3/2} \Upsilon \sqrt{x}+ \mathcal{O}(x) , \quad h=-\frac{1}{2}\ln{\frac{x}{\xi}} + \mathcal{O}\big(\sqrt{x}\big),  \label{k0met}
\end{align}
where $x \defeq r-r_\sg(t)$, and $\xi(t)$ is determined by the choice of time variable. The leading contributions to the reduced EMT components
\begin{align}
	& \tau_t \approx \tau^r = -\Upsilon^2 + \mathcal{O}\big(\sqrt{x}\big) , \label{eq:taut+r} \\
	& \tau_t^{~r} = \pm \Upsilon^2 + \mathcal{O}\big(\sqrt{x}\big) , \label{eq:taus}
\end{align}
are parametrized by $\Upsilon(t)>0$. The minus sign in Eq.~\eqref{eq:taut+r} is necessary to ensure that the solutions of the Einstein equations are real-valued \cite{bmmt:19,t:19}.

The near-horizon geometry is most conveniently expressed \cite{bmmt:19} in $(v,r)$ coordinates for $\tau_t^{~r}\approx - \Upsilon^2$, i.e.\ $r'_\sg<0$, and in $(u,r)$ coordinates for $\tau_t^{~r} \approx + \Upsilon^2$, i.e.\ $r'_\sg>0$. In both cases the metric functions are continuous across the horizons, and the expansions of ingoing and outgoing congruences can be readily evaluated. We see that the case $r_\sg'<0$ corresponds to an evaporating PBH, and $r_\sg'>0$ to an expanding white hole (contrary to erroneous interpretations of Refs.~\cite{t:19,t:20} that misidentified the latter as an accreting PBH). As our interest lies in the final stages of the collapse, we consider only evaporating PBH solutions in what follows.

Eq.~\eqref{gtr} must then hold identically, which yields the relation
\begin{align}
	r'_\sg/\sqrt{\xi}=-4\sqrt{\pi r_\sg}\,\Upsilon, 
	\label{rpr1}
\end{align}
where the prime denotes a derivative with respect to $t$. While the derivation uses the finiteness of  $T^\mu_{~\mu} = -R/8\pi$ and $T^{\mu\nu}T_{\mu\nu}=R^{\mu\nu}R_{\mu\nu}/64\pi^2$, all quadratic curvature invariants \cite{exact-e} are finite \cite{t:19,t:20}. This is also true for $k=1$ solutions that are described below.

For both black and white hole solutions the negative sign of $\tau_t$ and $\tau^r$ leads to the violation of the NEC \cite{mut:21,bmmt:19} in the vicinity of the apparent horizon. This can be deduced by studying a future-directed outward (inward) pointing radial null vector $k^\mu$ \cite{bmmt:19}.

Dynamical solutions with $k=1$ lead to finite energy density $\rho(t,r_\sg) \equiv E$ and pressure $p(t,r_\sg) \equiv P$. However, only their maximal possible values are consistent \cite{mut:21},
\begin{align}
	E=-P=1/\big(8\pi r_\sg^2\big),
\end{align}
and the corresponding metric functions are
\begin{align}
	C = r - c_{32}x^{3/2} + \mathcal{O}(x^2) , \quad h = - \frac{3}{2} \ln{\frac{x}{\xi}} + \mathcal{O}\big(\sqrt{x}\big) , \label{k1met}
\end{align}
where $c_{32}(t)>0$, and the consistency condition is
\begin{align}
	r'_\sg = - c_{32}\xi^{3/2} / r_\sg , \label{rgprime}
\end{align}
as we consider only evaporation.

Comparison of various expressions in $(t,r)$ and $(v,r)$ coordinates helps to establish many useful results. Since we use such comparisons quite extensively, we quote some useful expressions below. Components of the EMT are related by
 \begin{align}
	& \theta_v \defeq e^{-2h_+} \Theta_{vv} = \tau_t , \label{thev} \\
	& \theta_{vr} \defeq e^{-h_+} \Theta_{vr} = \big(\tau_t^{~r} - \tau_t\big)/f , \label{thevr} \\
	& \theta_r \defeq \Theta_{rr} = \big(\tau^r + \tau_t - 2 \tau_t^{~r}\big) / f^2 , \label{ther}
\end{align}
where $\Theta_{\mu\nu}$ is used to denote EMT components in $(v,r)$ coordinates.
The relevant Einstein equations then take the form
\begin{align}
	\partial_vC_+ &= 8 \pi e^{h_+} r^2 (\theta_v + \theta_{vr} f) , \\
	\partial_rC_+ &= - 8\pi r^2\theta_{vr} , \label{eq:vrEEvr} \\
	\partial_rh_+ &= 4 \pi r \theta_r .
\end{align}
An arbitrary spherically symmetric metric that is regular at the apparent horizon satisfies
\begin{align}
	& C_+(v,r) = r_+(v) + w_1(v) y + \mathcal{O}(y^2) , \label{cv1} \\
	& h_+(v,r) = \chi_1(v) y + \mathcal{O}(y^2) ,
\end{align}
where $y \defeq r - r_+(v)$, $w_1 \leqslant 1$, while $r_+(v) = r_\sg \big(t(v,r_+), r_\sg\big)$. The limits $\Theta^+_{\mu\nu} \defeq \lim_{r\to r_+} \Theta_{\mu\nu}$ yield
\begin{align}
	\theta_v^+ = (1 - w_1) \frac{r_+'}{8\pi r_+^2} , \quad \theta_{vr}^+ = - \frac{w_1}{8\pi r_+^2} , \quad \theta_r^+ = \frac{ \chi_1}{4\pi r_+} . \label{the3}
\end{align}
Both $k=0$ and $k=1$ solutions are needed to describe the formation of a black hole \cite{mut:21}. Assume that the first marginally trapped surface appears at some $v_\mathrm{S}$ at $r=r_+(v_\mathrm{S})$. For $v \leqslant v_\mathrm{S}$, the MS mass $C(v,r)/2$ in its vicinity is described in $(v,r)$ coordinates by
\begin{align}
	C_+(v,r) = \sigma(v) + r_*(v) + \sum_{i \geqslant 1}^\infty w_i(v)(r-r_*)^i,
\end{align}
where $r_*(v)$ corresponds to the maximum of $\Delta_v(r) \defeq C(v,r)-r$. The deficit $\sigma(v) \defeq \Delta_v\big(r_*(v)\big) \leqslant0$ by definition. At the advanced time $v_\mathrm{S}$ the location of the maximum corresponds to the first marginally trapped surface, $r_*(v_\mathrm{S}) = r_+(v_\mathrm{S})$, and $\sigma(v_\mathrm{S}) = 0$. For $v > v_\mathrm{S}$, the deficit $\sigma \equiv 0$ and the MS mass is described by Eq.~\eqref{cv1}.

For $v \leqslant v_\mathrm{S}$, the (local) maximum of  $\Delta_v$ satisfies $\partial \Delta_v / \partial r = 0$, hence $w_1(v) - 1 \equiv 0$. From Eqs.~\eqref{the3} and \eqref{thev} it follows that the newly formed black hole is described by a $k=1$ solution, since $w_1=1$ implies $\theta_v^+=0$ and thus $\Upsilon=0$. However, after its formation $r_+(v)$ is no longer a local maximum of $C_+(v,r)$, $w_1<1$, and thus at later times the black hole is described by a $k=0$ solution.

In the vicinity of the apparent horizon the equation for radial null geodesics becomes
\begin{align}
	\left. \frac{dr}{dt}\right|_{r=r_\sg} \hspace{-2mm} = \pm \left.e^hf\right|_{r=r_\sg} = \pm 4 \sqrt{\xi \pi r_\sg} \Upsilon = \mp{r'_\sg}, \label{genull}
\end{align}
where the upper (lower) signature corresponds to outgoing (ingoing) geodesics. This result indicates that massless particles cross the apparent horizon in finite time according to Bob. Massive particles likewise cross the apparent horizon in finite time $t$ \cite{t:19,bbgj:16}, unless they are too slow.

Some additional relations between the two sets of coordinates are useful: a point on the apparent horizon has the coordinates $\big(v,r_+(v)\big)$ and $\big(t,r_\sg(t)\big)$  in the two coordinate systems. Moving from $r_+(v)$ along the line of constant $v$ (i.e.\ along the ingoing radial null geodesic) by $\delta r$ leads to the point $(t+\delta t, r_\sg+\delta r)$. Using Eqs.~\eqref{rpr1} and \eqref{rgprime}, we obtain
\begin{align}
	\delta t=-\left.\frac{e^{-h}}{f}\right|_{r=r_\sg} \!\!\!\!\! \!\!\!\!\delta r=\frac{\delta r}{r'_\sg} + \mathcal{O}\big(\delta r^2\big) \label{dtrv}
\end{align}
for an evaporating black hole in both the $k=0$ and $k=1$ solutions. This implies that
\begin{align}
	\delta t( v, r_+ + y) = \frac{1}{r_\sg'} y + \frac{1}{2} \big( \partial_r^2 t \vert_{r=r_+} \big) y^2 + \mathcal{O}(y^3)
\end{align}
along the ingoing radial null geodesic, resulting in the relation
\begin{align}
	x(v,r_++y) &= r_+ + y-r_\sg\big(t(v,r_++y)\big) \nonumber \\
	& \hspace*{-13.3mm} = r_+ + y - \left[ r_\sg \big(t(v,r_+)\big) + r_\sg' \delta t + \tfrac{1}{2} r_\sg'' \delta t^2 + \mathcal{O}(\delta t^3) \right] \nonumber \\
	& \hspace*{-13.3mm} \eqdef \tfrac{1}{2} \omega^2 y^2
	\label{xyrel}
\end{align}
between the coordinates $x(t,r)$ and $y(v,r)$ in the vicinity of the apparent horizon, where 
\begin{align}
	\omega^2 = - r_\sg' \big(\partial_r^2 t \vert_{r=r_+}\big) - \frac{r_\sg''}{r_\sg'^2} ,
	\label{eq:omegaSquared}
\end{align}
and all derivatives are evaluated at $t=t(v,r_+)$. Using the invariance of the MS mass and expanding up to the first order in $y \defeq r-r_+(v)$, we obtain
\begin{align}
	C_+(v,r) &= r_+(v) + w_1 y + \ldots = C \big( t(v,r),r \big) \nonumber \\
	&= r_\sg \big( t(v,r_+) \big) + r_\sg' \bigg( \frac{y}{r_\sg'} \bigg) - 2 \sqrt{2 \pi r_\sg^3} \Upsilon \omega y + \ldots \nonumber \\
	&=r_++\left( 1 - 2 \sqrt{2 \pi r_\sg^3} \Upsilon \omega \right) y+\ldots,
\end{align}
and find 
\begin{align}
	w_1 = 1 - 2 \sqrt{2 \pi r_\sg^3} \Upsilon \omega \; \substack{{\scriptscriptstyle k=0}\\=} \; \frac{\sqrt{\pi}}{\Upsilon} r_\sg^{3/2} \left( e_{12} - p_{12} \right) , \label{w1t}
\end{align}
where the rightmost expression that is valid for $k=0$ solutions is obtained using Eq.~\eqref{eq:vrEEvr} and the limiting form of Eq.~\eqref{thevr} close to the horizon, and $e_{12}$ and $p_{12}$ denote the $\mathcal{O}(\sqrt{x})$ coefficients of the effective EMT components $\tau_t$ and $\tau^r$, respectively [cf.\  Eq.~\eqref{eq:effEMTtaus}].

\section{Parameter identification} \label{idUx}
The values of $\Upsilon$ and $\xi$ can be obtained from first principles only if one performs a complete analysis of the collapse of some matter distribution and the quantum excitations it generates. Such an analysis would provide a constructive proof of the existence of PBHs. In absence of such results we first obtain some general relations and then match them with the semiclassical results.

The apparent horizon of a PBH that was formed at a finite time of Bob is timelike \cite{bmmt:19}. Hence it is possible to introduce the induced metric
\begin{equation}
	ds^2|_{\mathrm{AH}}=-d\sigma^2+r_\ah d\Omega,
\end{equation}
where in the case of evaporation the proper time is most conveniently expressed in $(v,r)$ coordinates as $d\sigma = \sqrt{2|r_+'|}dv$. To remove the ambiguity we express coordinates of the apparent horizon as functions of proper time, such as $r_\ah(\sigma)$, $t_\ah(\sigma)$, and $v_\ah(\sigma)$. The invariance of the apparent horizon in spherically symmetric foliations means $r_\ah(\sigma) \equiv r_\sg\big(t_\ah(\sigma)\big)$, etc., and its rate of change is given by
\begin{align}
	\frac{dr_\ah}{d\sigma}=r'_\sg\big(t_\ah(\sigma)\big)\dot t_\ah=r'_+\big(v_\ah(\sigma)\big)\dot v_\ah. \label{ah-rate}
\end{align}
If one assumes that for an evaporating PBH $r_\sg$ is a monotonously decreasing function of time, one can write
\begin{align}
	\dot r_\ah=\Gamma_\ah(r_\ah), \quad r'_\sg=\Gamma_\sg(r_\sg), \quad r'_+=\Gamma_+(r_+),
\end{align}
where the relations between the functions $\Gamma_\ah$, $\Gamma_\sg$, and $\Gamma_+$ follow from Eq.~\eqref{ah-rate}. Without assuming any particular relation between $r_\sg'$ and $r_+'$, using Eq.~\eqref{thev} with $\tau_t = - \Upsilon^2 + \mathcal{O}(\sqrt{x})$ and comparing the first expression of Eq.~\eqref{the3} with Eq.~\eqref{w1t} leads to 
\begin{align}
	\Upsilon &= \frac{\sqrt{(1-w_1) \vert r_+' \vert}}{2 \sqrt{2 \pi} r_+} , \\
	\omega &= \sqrt{\frac{(1-w_1)}{r_+ \vert r_+' \vert}} ,
\end{align}
and from Eq.~\eqref{rpr1}, we obtain
\begin{align}
	\xi = \frac{r_\sg r_\sg'^2}{2 \vert r_+' \vert (1-w_1)} .
\end{align}
The semicalssical analysis is based on perturbative backreaction calculations that represent the metric as modified by the Hawking radiation that is produced by a slowly-varying sequence of Schwarzschild metrics. Identification of the Hawking temperature with the Kodama surface gravity (see Sec.~\ref{kappaT}) enforces $w_1=0$. This results in \cite{fn:book,bardeen:81,APT:19} $\Gamma_\sg(r)=\Gamma_+(r)$,
\begin{align}
	\frac{dr_\sg}{dt}=-\frac{\alpha}{r_\sg^2}, \qquad \frac{dr_+}{dv}=-\frac{\alpha}{r_+^2},  \label{paget}
\end{align}
where $\alpha$ denotes the emission rate coefficient. Using this result, we obtain
\begin{align}
	\Upsilon = \frac{\sqrt{\alpha}}{2 \sqrt{2 \pi} r_\sg^3} , \; \; \; \xi = \frac{\alpha}{2 r_\sg} , \; \; \; \omega = \sqrt{\frac{r_\sg}{\alpha}} ,
\end{align}
where the equalities on the rhs follow from Eq.~\eqref{paget}. We note  that this result agrees on the order of magnitude with the guess of Ref.~\cite{bmmt:19}, but as we will see below the assumptions of Ref.~\cite{bmt:19} are not fulfilled and its estimate is in general incorrect. If the $\partial^2_r t \vert_{r=r_+}$ term of Eq.~\eqref{eq:omegaSquared} is negligible, the evaporation time is radically different from the standard semiclassical results, namely
\begin{align}
	t_e \approx r_\sg (t_\mathrm{S}) \ln \frac{r_\sg(t_\mathrm{S})}{\beta} ,
\end{align}
for some positive coefficient $\beta$. The circumstances under which this is the case (if any) remain to be investigated.

\section{Temperature and surface gravity} \label{kappaT}
The surface gravity $\kappa$ plays an important role in GR, {particularly in black hole thermodynamics and more generally in }semiclassical gravity \cite{he:book,fn:book,faraoni:b}. For an observer at infinity the Hawking radiation that is produced on the background of a stationary black hole is thermal with its temperature given by $\kappa/2\pi$ \cite{fn:book,bmps:95}. However, surface gravity is unambiguously defined only in stationary spacetimes, where there are several equivalent definitions. These definitions are related to the inaffinity of null geodesics on the horizon, and to the peeling off properties of null geodesics near the horizon \cite{faraoni:b,kappa,kappaNielsenYoon}.

Stationary asymptotically flat spacetimes admit a Killing vector field $\xi^\mu$ that is timelike at infinity \cite{he:book,c:book,faraoni:b,exact-e}. A Killing horizon is a hypersurface on which the norm $\sqrt{\xi^\mu\xi_\mu}=0$. While logically this concept is independent of the notion of an event horizon, the two are related: for a black hole that is a solution of the Einstein equations in a stationary asymptotically flat spacetime the event horizon coincides with the Killing horizon \cite{fn:book,wald:01}.

A Killing orbit is the integral curve of the Killing vector field. The Killing property $\xi_{(\mu;\nu)}=0$ results in $\xi^\mu\xi_\mu=\mathrm{const}$ on each orbit. Coincidence of the two horizons allows one to introduce the surface gravity $\kappa$ as the inaffinity of null Killing geodesics on the event horizon,
\begin{align}
	\xi^\mu_{~;\nu}\xi^\nu \defeq \kappa \xi^\mu.
\end{align}
Assuming sufficient regularity of the metric, expansion of the null geodesics near the apparent horizon $r > r_\sg$ then establishes the concept of peeling affine gravity \cite{kappa,kappaNielsenYoon},
\begin{align}
	\frac{dr}{dt} = \pm 2 \kappa_\mathrm{peel}(t) x + \mathcal{O}(x^2). \label{peeld}
\end{align}
The two definitions coincide in stationary spacetimes. For a Schwarzschild metric with mass $M$ the surface gravity is $\kappa=1/(4M)=1/(2r_\sg)$.

Intuitively, the physical meaning of $\kappa$ can be interpreted as the force that would be required by an observer at infinity to hold a particle (of unit mass) stationary at the event horizon. Since the acceleration of a static observer will play a role in what follows, we reproduce here the derivation in $(t,r)$ coordinates. Consider an observer Eve at some fixed areal radius $r$. Her four-velocity is $u_\mathrm{E}^\mu = \delta^\mu_0 / \sqrt{-\sg_{00}}$, and her four-acceleration $a_\mathrm{E}^\mu=(0,\Gamma^r_{tt}/\sg_{00},0,0)$ in the Schwarzschild spacetime satisfies
\begin{align}
	g \defeq \sqrt{a^\mu_{\mathrm{E}} a_{{\mathrm{E}}\mu}} = \frac{r_\sg}{2r^2\sqrt{1-r_\sg/r}}.
\end{align}
Correcting by the redshift factor $z=-\sqrt{\sg_{00}}$ gives the surface gravity on approach to the horizon,
\begin{align}
	\kappa=\lim_{r\to r_\sg}zg=1/(2r_\sg).
\end{align}
Absence of the asymptotically timelike Killing vector in general dynamical spacetimes not only makes various analytic tasks computationally harder, but also requires generalization and reappraisal of the notions that are used in black hole physics.  Adapting one of the equivalent versions of surface gravity in stationary spacetimes is necessary.  For sufficiently slowly evolving horizons with properties sufficiently close to their classical counterparts these  different generalizations of surface gravity are practically indistinguishable \cite{kappa,kappaNielsenYoon}. This is important, as the role of the Hawking temperature is captured in various derivations either by the peeling \cite{peel:blsv} or the Kodama \cite{kpv:21} surface gravity. Indeed, gravitational collapse triggers radiation \cite{haj:87,blsv:06,vsk:07} that for macroscopic black holes at sufficiently late times approaches the standard Hawking radiation.

Nevertheless, this similarity fails for the self-consistent solutions that were described in Sec.~\ref{horizon}. Consider first the peeling surface gravity $\kappa_\mathrm{peel}$ \cite{mut:21}. For differentiable $C$ and $h$ the result is \cite{kappa,kappaNielsenYoon}
\begin{align}
	\kappa_\mathrm{peel} = \frac{e^{h(t,r_\sg)} \big( 1 - C'(t,r_\sg) \big)}{2 r_\sg} . \label{kap-peel}
\end{align}
However, such an expansion is impossible for both $k=0$ and $k=1$ solutions. The metric functions of Eqs.~\eqref{k0met} and \eqref{k1met} lead to a divergent peeling gravity. This happens because Eq.~\eqref{genull} ensures that there is a nonzero constant term in the expansion of the geodesics, and instead of Eq.~\eqref{peeld} we have
\begin{align}
	\frac{dr}{dt} = \pm r_\sg'+a_{12}(t)\sqrt{x} + \mathcal{O}(x) ,
\end{align}
where $a_{12}$ depends on the higher-order terms of the EMT. Similarly, the redshifted acceleration of a static observer diverges as
\begin{align}
	zg = \frac{|r'_\sg|}{4x} + \mathcal{O}(x^{-1/2}) .
\end{align}
However, the peeling surface gravity was originally introduced using regular Painlev\'{e}--Gullstrand coordinates $(\bt,r)$ \cite{nv:06} (whose properties are briefly summarized in App.~\ref{a:pg}). In fact, the two possible definitions are \cite{nv:06}
\begin{align}
	\kappa_{\PG_1}=\left.\frac{1}{2r_\sg}(1-\pad_r \bar{ C})\right|_{r=r_\sg}, \label{kpg1}
\end{align}
where $\bC=C\big(t(\bt,r),r\big)$ is the MS mass in Painlev\'{e}--Gullstrand coordinates, and \cite{vad:11}
\begin{align}
	\kappa_{\PG_2}=\left.\frac{1}{2r_\sg}(1-\pad_r \bar{ C}+\pad_\bt \bC)\right|_{r=r_\sg}. \label{kpg2}
\end{align}
Using the invariance of the MS mass, we have
\begin{align}
	\frac{\pad \bar C}{\pad r}=\left.\frac{\pad  C}{\pad t}\frac{\pad t}{\pad r}\right|_{\bt}+\frac{\pad C}{\pad r}.
\end{align}
Recalling that for an evaporating PBH
\begin{align}
	\lim_{r\to r_\sg} f(t,r)e^{h(t,r)}=-r'_\sg,
\end{align}
we have [selecting the positive sign in Eq.~\eqref{pde1}]
\begin{align}
	\left.\frac{\pad t}{\pad r}\right|_{\bt}=-\frac{{\pad \bt}/{\pad r}}{{\pad\bt}/{\pad t}}\to \frac{1}{r'_\sg}.
\end{align}
For $k=0$ solutions, we then have for $r \to r_\sg$
\begin{align}
	&\frac{\pad C(t,r)}{\pad t}=r'_\sg\left(1+\frac{2\sqrt{\pi r_\sg^3}\Upsilon}{\sqrt{r-r_\sg}}\right)+\mathcal{O}\big(\sqrt{x}\big), \label{dtC} \\
	&\frac{\pad C(t,r)}{\pad r}=-\frac{2\sqrt{\pi r_\sg^3}\Upsilon}{\sqrt{r-r_\sg}}+\mathcal{O}\big(\sqrt{x}\big).
\end{align}
Substituting everything into the definition Eq.~\eqref{kpg1} results in
\begin{align}
	\kappa_{\PG_1}=0.
\end{align}
Furthermore, we also obtain
\begin{align}
	\kappa_{\PG_2}=\left.\frac{ \pad_\bt \bC}{2r_\sg}\right|_{r=r_\sg}.
\end{align}
Since
\begin{align}
	\pad_\bt\bC=\pad_t C\pad_\bt t|_r,
\end{align}
we find using Eq.~\eqref{dtC} that
\begin{align}
	\pad_\bt \bC\approx\frac{r'_\sg}{\pad_t\bt}\left(1+\frac{2\sqrt{\pi r_\sg^3}\Upsilon}{\sqrt{r-r_\sg}}\right),\label{mess1}
\end{align}
which in the limit $r \to r_\sg$ results in three distinct possibilities that depend on the behavior of the function $\bt(t,r)$. If as $r\to r_\sg$ the Painlev\'{e}--Gullstrand time $\bt$ diverges faster than $1/\sqrt{r-r_\sg}$, then  $ \kappa_{\PG_2}=0=\kappa_{\PG_1}$. If  $\bt$ diverges slower than $1/\sqrt{r-r_\sg}$, then   $ \kappa_{\PG_2}$ is divergent. Finally,
\begin{align}
	\bt=\tau(t)\sqrt{r-r_\sg}+\mathcal{O}(r-r_\sg),
\end{align}
where $\tau(t)$ is some function, leads to a finite value of $\kappa_{\PG_2}$. In fact, this form is consistent with the limiting form of Eq.~\eqref{pde1} (see App.~\ref{a:pg} for details).

The Kodama vector field can be introduced in any spherically symmetric spacetime \cite{k:80,av:10}. It has many useful properties of the Killing field to which, modulo possible rescaling, it reduces in the static case \cite{faraoni:b,kappa,kappaNielsenYoon,av:10}. Similar to the Killing vector, it is most conveniently expressed in $(v,r)$ coordinates,
\begin{align}
	K^\mu = \big(e^{-h_+},0,0,0\big).
\end{align}
It is covariantly conserved, and generates the conserved current
\begin{align}
	& \nabla_\mu K^\mu = 0 , \\
	& \nabla_\mu J^\mu = 0 , \quad J^\mu \defeq G^{\mu\nu} K_\nu ,
\end{align}
where $G_{\mu\nu}=R_{\mu\nu} - \tfrac{1}{2} \sg_{\mu\nu}R$ is the Einstein tensor, {thereby giving} a natural geometric meaning to the Schwarzschild coordinate time $t$. The MS mass is its Noether charge.

Since $K_{(\mu;\nu)}\neq 0$, the generalized Hayward--Kodama surface gravity is defined via \cite{hay:98}
\begin{align}
	\frac{1}{2} K^\mu(\nabla_\mu K_\nu-\nabla_\nu K_\mu) \defeq \kappa_\mathrm{K} K_\nu,
\end{align}
evaluated on the apparent horizon. Hence
\begin{align}
	\kappa_\mathrm{K} = \frac{1}{2} \left.\left(\frac{C_+(v,r)}{r^2} - \frac{\partial_r C_+(v,r)}{r}\right) \right|_{r=r_+} \hspace*{-3mm} = \frac{(1-w_1)}{2r_+}, \label{eq:surfgKodama}
\end{align}
where we used Eq.~\eqref{cv1} to obtain the final result. Thus at the formation of a black hole (i.e.\ of the first trapped surface) this version of surface gravity is zero. At the subsequent evolution stages that correspond to a $k=0$ solution, $\kappa_\mathrm{K}$ is nonzero. However, it approaches the static value $\kappa=1/(4M)$ only if the metric is close to the pure Vaidya metric with $w_1\equiv 0$.

\section{Discussion}
Our considerations have shown that a proper formulation of the information loss paradox is quite subtle, and that its standard exposition at the very least warrants considerable revision. Formation of the apparent horizon at some finite time $t_\mathrm{S}$ that distant Bob measures is a necessary condition to set up the information loss problem. A consistent solution of the field equations admits evaporation whilst yielding regularity at the horizon, but necessarily entails a violation of the NEC.

The necessity of the NEC violation is an obstacle, as it requires a mechanism to convert the original collapsing matter into exotic matter that must be present in the vicinity of the forming apparent horizon. Conventional mechanisms for mass loss, such as the emission of gravitational waves, should work in tandem with production of the negative-energy-density matter. Collapse-induced Hawking-like radiation is thus not only a necessary quantum-mechanical ingredient of the paradox, but is necessary for producing its classical setting.

This brings us to a more serious difficulty: two ``close'' generalizations of surface gravity [namely the peeling surface gravity \eqref{kap-peel} and the Hayward--Kodama surface gravity \eqref{eq:surfgKodama}] that underpin different derivations of Hawking radiation on the background of an evolving spacetime are irreconcilable. In fact, three versions of the same peeling surface gravity [Eqs.~\eqref{kap-peel}, \eqref{kpg1}, and \eqref{kpg2}] are irreconcilable as well. Moreover, it is not clear if the required structure of the EMT can be matched \cite{mut:21}.

In addition, if the Hawking temperature is indeed proportional to the peeling surface gravity, then black holes explode (or freeze) on their formation. In this case the semiclassical picture is not valid, and it is impossible to formulate the information loss problem. Alternatively, if the Hawking temperature is proportional to the Kodama surface gravity, then it vanishes at the formation of a black hole; although it increases during evaporation, it should reach zero again at the final stages of the evaporation process \cite{dst:22}. If the Kodama surface gravity reaches the classical value $\kappa_\mathrm{K}=1/(2r_\sg)$, then it cannot be the black hole temperature. Moreover, it is not clear how, given indications to the contrary \cite{C-R:18}, a process with close to zero flux can ensure the necessary dominance of quantum effects over normal matter in the vicinity of the outer apparent horizon.

Our analysis indicates that the circumstances surrounding the formation of PBHs do not provide a basis to formulate the information loss problem within the semiclassical framework. Therefore, in order to resolve the ``paradox'', new physics is required to provide a mechanism to explain why information is lost to begin with, and describe how this process may occur in a self-consistent way. It should be noted that, even if the issues that have been raised so far are resolved, scrutiny of the precise technical aspects of commonly invoked semiclassical notions indicate that ``Page time unitarity'' may appear to be violated even if the underlying physics is unitary \cite{SFA:21}. A recent study \cite{bfm:21} that is complementary to the argumentation presented here also indicates that the standard form of the paradox can be consistently rendered only if new physics begins to play a role before reaching the Planck scale.

\acknowledgments
RBM is supported by the Natural Sciences and Engineering Research Council of Canada. SM is supported by an International Macquarie University Research Excellence Scholarship and a Sydney Quantum Academy Scholarship. The work of DRT is supported by the ARC Discovery project grant No.\ DP210101279.

\appendix
\section{Painlev\'{e}--Gullstrand coordinates} \label{a:pg}
One possible set of coordinates that are regular across the horizon \cite{nv:06,vad:11,mp:01} is obtained by taking the proper time of an infalling observer (with zero initial velocity at infinity) as the time coordinate. The Painlev\'{e}--Gullstrand time $\bar{t}$ for the Schwarzschild metric is given by \cite{mp:01}
\begin{align}
	\bt=t+2\sqrt{r_\sg r}+r_\sg\ln\left|\frac{\sqrt{r}-\sqrt{r_\sg}}{\sqrt{r}+\sqrt{r_\sg}}\right|,
\end{align}
and the metric takes the form
\begin{align}
	ds^2=-fd\bt^2+2\sqrt{r_\sg/r}d\bt dr+dr^2+r^2d\Omega. \label{pgs}
\end{align}
For a general metric of Eq.~\eqref{sgenm}, writing \cite{nv:06}
\begin{align}
	d\bt=\pad_t\bt dt+  {\pad_r \bt dr}, \label{dtbar}
\end{align}
and requiring that in these coordinates, similar to Eq.~\eqref{pgs}, the metric component $\sg'_{rr}=1$, leads to the first-order linear homogenous partial differential equation for the Painlev\'{e}--Gullstrand time
\begin{align}
	\pad_r\bt=\pm \sqrt{\frac{C}{r}}\frac{e^{-h}}{f}\pad_t\bt. \label{pde1}
\end{align}
Subject to appropriate boundary conditions this equation has a unique solution. The metric in $(\bt,r)$ coordinates is then
\begin{align}
	ds^2= -\frac{e^{2h}}{(\pad_t \bt)^2}d\bt^2 \pm 2 \frac{e^h}{\pad_t\bt}\sqrt{\frac{C}{r}}d\bt dr+dr^2+r^2d\Omega,  \label{pgg}
\end{align}
where the metric functions are $C\big(t(\bt,r),r\big) \eqdef \bar C(\bt,r)$ and $h\big(t(\bt,r),r\big) \eqdef \bar h(\bt,r)$. To match the Painlev\'{e}--Gullstrand  coordinates for the Schwarzschild spacetime we select the upper sign  in the above expressions.

We now identify the scaling of $\pad_t \bt$ on the apparent horizon by considering the changes in $t_\ah(\sigma)$ and $\bt_\ah(\sigma)$,
\begin{align}
	\frac{\dot{\bt}_\ah}{\dot{t}_\ah}=\left.\frac{d\bt}{dt}\right|_\mathrm{AH}\equiv\frac{d\bt\big(t,r_\sg(t)\big)}{dt}.
\end{align}
As $r\to r_\sg$
\begin{align}
	\sqrt{\frac{C}{r}}\frac{e^{-h}}{f}=\frac{1}{|r_\sg'|}+\Delta(t)\sqrt{r-r_\sg}+\mathcal{O}(r-r_\sg),
\end{align}
where $\Delta(t)$ depends on the higher-order terms in the metric.
  Taking into account Eqs.~\eqref{dtbar} and \eqref{pde1}, we have
\begin{align}
	\left.\frac{d\bt}{dt}\right|_\mathrm{AH}&=\lim_{r\to r_\sg}\frac{\pad \bt}{\pad t}\left(1+ \sqrt{\frac{C}{r}}\frac{e^{-h}}{f}\frac{dr(t)}{dt}\right)\\
&=\lim_{r\to r_\sg}\pad_t\bt\, r_\sg' \,\Delta \sqrt{r-r_\sg }.
\end{align}
As both $t$ and $\bt$ are finite throughout the evolution of the apparent horizon, we must conclude that $\pad_t\bt$ diverges as
\begin{align}
	\left.\frac{\pad \bt}{\pad t}\right|_{r\to r_\sg}\propto\frac{1}{\sqrt{r-r_\sg}}.
\end{align}
Then Eq.~\eqref{mess1} implies that the surface gravity $\kappa_{\PG_2}$ is finite,
\begin{align}
	\kappa_{\PG_2}=2\sqrt{\pi r_\sg^3}\Upsilon\Delta.
\end{align}

\end{document}